\RequirePackage{ifpdf}
\documentclass[12pt]{article}
\usepackage{epsfig,graphicx,bm,amsmath,amsfonts}
\usepackage{epstopdf}
\usepackage{amssymb,xcolor,float}
\usepackage{amsthm}
\usepackage{hyperref}
\usepackage{cite}
\input epsf.sty
\usepackage{cite}
\usepackage{comment}
\usepackage{epstopdf}
\usepackage{float}
\usepackage{caption}
\usepackage{subcaption}
\usepackage{enumitem}
\usepackage{bbm}
\usepackage{bm}
\usepackage{longtable}
\usepackage{parskip}
\usepackage{enumitem}
\usepackage{graphicx,epsfig}
\usepackage{ae,aecompl}
\usepackage{dcolumn}
\usepackage{latexsym}
\usepackage{psfrag}
\usepackage{ytableau}
\usepackage{tabularx}
\usepackage{float}
\usepackage{wrapfig}

\usepackage{xcolor}
\usepackage[english]{babel}
\usepackage[T1]{fontenc}
\usepackage[utf8]{inputenc}
\usepackage{authblk}
\usepackage{mathtools}
\usepackage{slashed}
\usepackage{mattens,amssymb}
\usepackage{amsmath,amssymb}
\usepackage{amsfonts}
\usepackage{graphicx,color}
\usepackage{cite}
\usepackage{hyperref}
\hypersetup{
	colorlinks=true,
	linkcolor=blue,
	filecolor=magenta,      
	urlcolor=cyan,
}
\usepackage[capitalize]{cleveref}
\numberwithin{equation}{section} 

\definecolor{refcol}{rgb}{0.9,0.1,0.1}
\hypersetup{colorlinks=true,linkcolor=blue,citecolor=refcol,urlcolor=cyan,linktocpage}

\usepackage{wrapfig}

\newcommand{\Tr}{\text{Tr}}

\newcommand{\ben}{\begin{eqnarray}\displaystyle}
\newcommand{\een}{\end{eqnarray}}

\newcommand{\be}{\begin{equation}}
\newcommand{\ee}{\end{equation}}

\newcommand{\bc}{\begin{center}}
\newcommand{\ec}{\end{center}}

\newcommand{\eesp}{\end{split}}
\newcommand{\bsp}{\begin{split}}

\newcommand{\Rmnum}[1]{\expandafter\@slowromancap\romannumeral #1@}
\newcommand{\cB}{\mathcal{B}}

\newcommand{\cH}{\mathcal{H}}

\newcommand{\lb}{\left (}
\newcommand{\rb}{\right )}

\def\tL{\text{L}}

\newcommand{\where}{\text{where}}

\newcommand{\tand}{\text{and}}

\newcommand{\bensp}{\begin{eqnarray}\begin{split}}
\newcommand{\eensp}{\end{eqnarray}\end{split}}

\newcommand{\bnm}{\begin{matrix}}
\newcommand{\enm}{\end{matrix}}
\def\XXint#1#2#3{{\setbox0=\hbox{$#1{#2#3}{\int}$ }
\vcenter{\hbox{$#2#3$ }}\kern-.6\wd0}}

\newcommand{\ket}[1]{|#1\big>}
\newcommand{\bra}[1]{\big<#1|}

\newcommand{\braket}[2]{\big<#1|#2\big>}

\begin{document}

\begin{center}
{\Large \bf{Bosonisation and BTZ Black Hole Microstates}}
\end{center}
\baselineskip=18pt

\bigskip

\centerline{Suvankar Dutta$^{a}$, Shruti Menon$^{a}$ and Aayush Srivastav$^{b}$}

\bigskip

\centerline{\large \it $^{a}$Indian Institute of Science Education and Research Bhopal}
\centerline{\large \it Bhopal bypass, Bhopal 462066, India}
\centerline{\large \it $^{b}$Indian Institute of Science Education and Research Kolkata }
\centerline{\large \it Mohanpur, Nadia 741246, India}

\bigskip

\centerline{E-mail: suvankar@iiserb.ac.in, shruti98.menon@gmail.com, aayush1309physics@gmail.com}

\vskip .6cm
\medskip

\vspace*{4.0ex}

\begin{abstract}

When the boundary dynamics of \(AdS_3\) gravity is governed by the collective field theory Hamiltonian proposed by Jevicki and Sakita, its asymptotic symmetry algebra becomes the centerless \(U(1)\) Kac-Moody algebra. We quantize this system using the quantum bosonization of relativistic free fermions and relate these to the dynamical fields of \(AdS_3\) gravity. This leads to a correspondence where different bulk configurations correspond to distinct states (particle-hole pair excitations) in the fermionic Hilbert space. This mapping allows us to construct BTZ black hole microstates, represented by Young diagrams of irreducible \(U(\infty)\) representations. Notably, the logarithm of the microstate degeneracy exactly reproduces the classical entropy of the BTZ black hole.

\end{abstract}

\vfill\eject

\tableofcontents

\section{Introduction}

Three-dimensional spacetime, whether it incorporates a cosmological constant or not, serves as a powerful and precise framework for investigating the intricate classical and quantum facets of gravity. The compelling reason for this is that, in three dimensions, the dynamics of classical gravity are devoid of bulk degrees of freedom; the asymptotic boundary conditions entirely dictate it. Notably, in the context of a negative cosmological constant, Brown and Henneaux revealed that, with the right choice of boundary conditions, the asymptotic symmetry group can be elegantly generated by two copies of the Virasoro algebra \cite{Brown:1986nw}. While the boundary conditions proposed by Brown and Henneaux are considered to be the standard one, it is essential to recognize that other boundary conditions have also been studied rigorously \cite{Henneaux:2013dra, Bunster:2014mua, Perez:2016vqo,Perez:2012cf,Afshar:2016wfy,Grumiller:2016pqb,Grumiller:2016kcp,Grumiller:2019tyl,Afshar:2016kjj,Afshar:2013vka, Troessaert:2013fma, Avery:2013dja, Ammon:2017vwt,Ozer:2019nkv, Gonzalez:2018jgp, Campoleoni:2010zq}. Further exploring this broader perspective can deepen our understanding of the fundamental nature of gravity.

In $AdS_3$, gravity allows for a new family of non-standard boundary conditions, where the chemical potentials (the time component of the gauge field) are no longer fixed at the asymptotic boundary. Instead, they can explicitly depend on the fields (specifically, the angular components of the gauge fields), and the relationship between them varies according to the choice of boundary Hamiltonians. Consequently, the classical dynamics of boundary gravitons are governed by the Hamiltonian. Following the work of \cite{Dutta:2023uxe}, in this paper, we consider the boundary Hamiltonian to be the collective field theory Hamiltonian proposed by Jevicki and Sakita \cite{sakita, jevicki}. We refer to this boundary condition as the collective field theory (ColFT) boundary condition. In our approach, the ColFT Hamiltonian is defined over a cylinder and can be derived from unitary matrix quantum mechanics in the presence of an arbitrary potential $W(\Tr U)$. These boundary conditions permit a BTZ black hole solution in the bulk. The equations of motion for the collective field and its conjugate momentum are analogous to the continuity and Euler's equations of a one-dimensional fluid moving on a circle under negative pressure, with the fluid density represented by the collective field. This setup allows us to conceptualize the fluid as residing on the boundary of the \( AdS_3 \) spacetime. Therefore, the dynamics of the boundary gravitons are described by the dynamics of this boundary one-dimensional fluid. We find that different configurations of the boundary fluid correspond to various bulk geometries. Specifically, the BTZ black hole (a time-independent spherically symmetric solution) corresponds to a fluid with constant density and velocity. An extremal BTZ black hole corresponds to a scenario where the fluid's velocity equals its density, while a non-rotating BTZ black hole geometry is associated with a static fluid.

We quantize this system by promoting the classical Poisson structure to commutator brackets. This leads us to discover that the asymptotic symmetry algebra is isomorphic to the \( U(1) \) Kac-Moody algebra without any central charge. To find the representation of this algebra, we utilize the quantum bosonization of relativistic free fermions and connect them to the dynamical fields of \( AdS_3 \) gravity. This approach allows us to construct the Hilbert space of the free fermionic system and map bulk configurations (bosonic excitations) to various particle-hole pair states in the fermionic Hilbert space. As a result, we construct the microstates of the BTZ black hole for a given mass and angular momentum. We find that these different microstates can be represented by Young diagrams of irreducible \( U(\infty) \) representations with a given number of boxes, where the number of boxes depends on the mass and angular momentum of the black hole. Finally, we compute the degeneracy of these microstates and observe that the logarithm of the degeneracy exactly matches the classical entropy of the BTZ black hole.

The structure of the paper is outlined as follows. We briefly discuss $AdS_3$ gravity and the non-standard boundary conditions in Section \ref{sec:Ads3gravity}. In the subsequent section (Section \ref{sec:colHamBTZ}), we demonstrate how to derive the BTZ black hole geometry for ColFT boundary conditions. In Section \ref{sec:quantisation}, we focus on quantizing the system. This section discusses the bosonization of relativistic fermions, followed by the construction of the Hilbert space. We then identify various states within the Hilbert space corresponding to a given black hole geometry. Finally, we conclude the paper with a summary of our results and some intriguing related questions.

\section{Gravity in $AdS_3$ and Chern-Simons theory}
\label{sec:Ads3gravity}

This section provides a concise overview of the relationship between three-dimensional Anti-de Sitter spacetime and Chern-Simons theory. The theory of gravity in the bulk can be formulated using Chern-Simons theory with the gauge group \( SO(2,2) \) \cite{Campoleoni:2010zq, Campoleoni:2011hg}. Since the representation of \( so(2,2) \) is isomorphic to \( sl(2,\mathbb{R}) \oplus sl(2,\mathbb{R}) \), the Chern-Simons gauge fields can be decomposed into two chiral sectors, which we denote as \( A^\pm \). In each sector, the matrix-valued gauge fields \( A^\pm \) can be expressed in terms of the vielbein \( e^a_\mu \) and the spin connection \( \omega^{ab}_{\mu} \)
\begin{align}\label{eq:gaugemetricrel}
A^\pm = \left(\omega^a \pm \frac{e^a}{l}\right) T_a,
\end{align}
where \( l \) is the radius of the \( AdS_3 \) space and \( T_a \) are the generators of \( SL(2,\mathbb{R}) \). The quantities \(\omega_a\) are dual to the spin connection and can be related to \(\omega^{bc}\) by the following equation
\begin{equation}
\omega_a \equiv \frac{1}{2} \epsilon_{abc} \omega^{bc}.
\end{equation}
Using the relationship between the vielbein, the spin connection, and gauge fields, it can be demonstrated that the Einstein-Hilbert action is equivalent to the Chern-Simons (CS) action. This can be expressed as
\begin{align}\label{eq:CSacn}
I = I_{CS}(A^+) - I_{CS}(A^-)
\end{align}
where the Chern-Simons action for \(A^\pm\) is given by
\begin{align}\label{eq:CSacn2}
I_{CS}(A^\pm) = \frac{\mathrm{k}}{4\pi} \int \Tr\left[ A^\pm \wedge dA^\pm + \frac{2}{3} (A^\pm)^3 \right] + \cB_\infty(A^\pm).
\end{align}
Here, \(\mathrm{k}\) represents the level of the Chern-Simons theory. It is related to the three-dimensional Newton's constant \(G\) and the radius \(l\) of \(AdS_3\) space through the equation
\begin{equation}\label{eq:kGrelation}
\mathrm{k} = \frac{l}{4G}.
\end{equation}

In \eqref{eq:CSacn2}, $\cB_\infty(A^\pm)$ is a boundary term added to the bulk action to make $\delta I_{CS}(A^\pm) = 0$. The trace in (\ref{eq:CSacn2}) acts on the generators of the algebra $\mathbf{sl}(2,R)$ in the fundamental representation. The generators of $\mathbf{sl}(2,R)$ are given by $\tL_\pm$ and $\tL_0$. Trace acts on the generators $\tL_0$ and $\tL_\pm$ as,
\begin{align}
     \Tr( \tL_0 \tL_0) = \frac{1}{2}, \quad \Tr( \tL_1 \tL_{-1}) = -1
\end{align}
and the rest of the combinations are zero.

Using the relationship between the metric and the vielbein, we can compute the different components of the metric from the gauge fields
\begin{align}\label{eq:metric}
g_{\mu\nu} = \frac{l^2}{2} \Tr((A^+ - A^-)_\mu (A^+ - A^-)_\nu).
\end{align}

In three dimensions, gravity is locally trivial; therefore, all dynamics occur near the boundary and are sensitive to the boundary conditions.

\subsection{Boundary conditions}

We define the coordinates of the Lorentzian \( AdS_3 \) manifold as $r$, $t$, and $\theta$. Here,  $\theta$ is a compact coordinate, $r$ represents the radial direction, with the asymptotic region (boundary) located at $r \to \infty$, and $t$ denotes time. We adopt the following form for the gauge fields
\begin{equation}\label{eq:Apm-apm}
A^\pm = b_\pm^{-1} \lb d + a^\pm \rb b_\pm
\end{equation}
where $b^\pm$ are gauge group elements that depend only on the radial coordinate $r$. The connections $a^\pm$ depend solely on the transverse coordinates $t$ and  \(\theta\). To satisfy Maxwell's equations, which are equivalent to Einstein's equations when expressed in terms of the metric, it is not necessary to specify a particular form for \( b_\pm \). We choose the connection in the following form, as discussed in \cite{Grumiller:2016kcp}
\begin{equation} \label{eq:apmform}
a^\pm(t,\theta) = \left( \xi_\pm(t,\theta) dt \pm p_\pm(t,\theta) d\theta \right) \tL_0.
\end{equation}
Here, \( p_\pm \) are the dynamical fields associated with gravity, while \( \xi_\pm \) correspond to the chemical potentials. The Maxwell equations derived from the action are given by
\begin{eqnarray}\label{eq:Maxeq}
    dA^\pm + A^{\pm 2} =0.
\end{eqnarray}
By substituting the forms of \( A^\pm \) and \( a^\pm \), we can verify that the dynamical fields and the chemical potentials satisfy the following relation
\be\label{eq:maxeq}
\Dot{p}_\pm(t,\theta) = \pm \xi_\pm'(t,\theta).
\ee
In this equation, \( \cdot \) and \( ' \) denote partial derivatives with respect to \( t \) and \( \theta \), respectively.

To determine the boundary term \(\cB_\infty^\pm\), we require that \(\delta I_{CS} = 0\) for any arbitrary variations of the gauge fields. This leads to the following equation
\begin{eqnarray}\label{eq:Bterms}
    \delta \cB_\infty^\pm = - \frac{\mathrm{k}}{2\pi} \int dt \, d\theta \, \langle A_t^\pm \delta A^\pm_{\theta} \rangle = \mp \frac{\mathrm{k}}{4\pi} \int dt \, d\theta \, \xi_\pm \delta p_\pm.
\end{eqnarray}
If \(\xi_\pm\) is constant, it is possible to factor \(\delta\) out of the integral to compute the boundary term. However, this situation is somewhat exceptional. In the more general case, we define the chemical potential \(\xi_\pm\) in terms of the variation of a quantity \(H^\pm\) with respect to \(p_\pm\)
\begin{eqnarray}\label{eq:xidef}
    \xi_\pm = -\frac{4\pi}{\mathrm{k}} \frac{\delta H^\pm}{\delta p_\pm},
\end{eqnarray}
where \(H^\pm\) is typically a functional of \(p_\pm\) and its various derivatives with respect to \(\theta\)
\begin{equation}
H^\pm = \int d\theta \ \cH^\pm(p_\pm, p_\pm', \cdots).
\end{equation}
Consequently, the boundary term can be expressed as
\begin{eqnarray}\label{eq:boundterm}
    \cB_\infty^\pm = \pm \int dt \, d\theta \, \cH^\pm = \pm \int dt \, H^\pm.
\end{eqnarray}

The dynamics of the fields \( p_\pm \) at the boundary depend on the choice of \( \mathcal{H}^\pm \). The equations governing these dynamics are given by
\begin{eqnarray}\label{eq:boundarydynamics}
   \Dot{p}_\pm(t, \theta) = \mp \frac{4\pi}{\mathrm{k}} \frac{\partial }{\partial \theta} \lb  \frac{\delta H^\pm}{\delta p_\pm} \rb.
\end{eqnarray}
We refer to \( H^\pm \) as the boundary Hamiltonian. Different choices of boundary Hamiltonian lead to distinct boundary dynamics. The equations of motion presented above can be derived from the following Poisson structure
\begin{equation}
    \dot p_\pm(t,\theta) = \{p_{\pm}(t,\theta),H^\pm\}_{PB}
\end{equation}
where the Poisson bracket is defined as
\begin{equation}\label{eq:poisson}
    \{p_{\pm}(t,\theta), p_{\pm}(t,\theta')\}_{PB} = \mp \frac{4\pi}{\mathrm{k}} \frac{\partial}{\partial\theta} \delta(\theta-\theta').
\end{equation}

\section{Collective field theory and black holes in $AdS_3$ gravity}
\label{sec:colHamBTZ}

A new set of boundary conditions for \(AdS_3\) gravity was proposed in \cite{Dutta:2023uxe}, where the dynamics at the boundary for spin-two and higher-spin fields are described by the interacting collective field theory Hamiltonian developed by Jevicki and Sakita.

The dynamics of the collective field and its conjugate momentum resemble the evolution of a one-dimensional fluid with negative pressure. By selecting the boundary Hamiltonian as a collective field theory Hamiltonian, we can map various bulk geometries to one-dimensional fluid motions. Our focus is on time-independent, spherically symmetric black hole solutions. We find that black holes with specified mass and angular momentum are equivalent to a one-dimensional fluid with constant density moving in a circular path at a constant velocity.

The Hamiltonian for the collective field theory is given by
\begin{equation}
    H_{CFT} = \int d\theta \ \sigma(t,\theta) \left[ \frac{1}{2} \left( \frac{\partial \Pi(t,\theta)}{\partial \theta}\right)^2 + \frac{\pi^2}{6} \sigma^2(t,\theta) + W(\theta) \right]
\end{equation}
where \(\sigma(t,\theta)\) represents the eigenvalue density of a \(N \times N\) unitary matrix \(U\) in the continuum (large \(N\)) limit, and \(\Pi(t,\theta)\) is the momentum conjugate to \(\sigma(t,\theta)\). The following equations describe the dynamics of the collective field and its conjugate momentum
\begin{eqnarray}
    \begin{split}
        \partial_t \sigma(t,\theta) + \partial_\theta (\sigma(t,\theta) v(t,\theta)) & = 0, \\
        \partial_t v(t,\theta) + v(t,\theta)\partial_\theta  v(t,\theta)  & = - \partial_\theta \left(\frac{\pi^2}{2} \sigma^2(t,\theta) + W(\theta)\right).
    \end{split}
\end{eqnarray}
These equations represent the continuity equation and Euler's equation, respectively, for a one-dimensional fluid moving on a circle with negative pressure and velocity given by
\begin{equation}
    v(t,\theta) = \partial_\theta \Pi(t,\theta).
\end{equation}
In this context, $\sigma$ represents the fluid density. The above set of equations is a coupled system of non-linear partial differential equations. To decouple these equations, we can introduce two new variables, $p_\pm(t,\theta)$, defined as
\begin{equation}\label{eq:ppmsv}
    p_\pm(t,\theta) = v(t,\theta) \pm \pi \sigma(t,\theta).
\end{equation}
The equations of motion are then simplified to
\begin{equation}
    \partial_t p_\pm(t,\theta) + p_\pm(t,\theta) \partial_\theta p_\pm(t,\theta) + W'(\theta) = 0
\end{equation}
and the Hamiltonian can be expressed as
\begin{equation}
    H_{CFT} = H_{CFT}^+ + H_{CFT}^-, \quad \where \quad H_{CFT}^\pm = \pm \frac{1}{2\pi} \int d\theta \left(\frac{p^3_\pm(t,\theta)}{6} + W(\theta) p_\pm(t,\theta) \right).
\end{equation}

In this study, we examine the boundary dynamics (\ref{eq:boundarydynamics}) governed by the ColFT Hamiltonian
\begin{equation}\label{eq:HamCFT}
	H^\pm = \frac{\mathrm{k}}{2} H_{CFT}^\pm = \pm \frac{\mathrm{k}}{4\pi} \int d\theta \left(\frac{p^3_\pm(t,\theta)}{6} + W(\theta) p_\pm(t,\theta) \right).
\end{equation}

\subsection{Asymptotic symmetries and conserved charges}

The asymptotic symmetries are defined by a set of gauge transformations that maintain the asymptotic form of the gauge fields. In this context, the asymptotic form of the gauge field, as specified in equation (\ref{eq:apmform}), is preserved under the gauge transformation given by
\begin{equation}\label{eq:gaugetrana}
    \delta a^\pm = d\lambda^\pm + [a^\pm, \lambda^\pm].
\end{equation}
Here, we choose the gauge transformation parameter as \(\lambda^\pm = \eta^\pm(t,\theta)\tL_0\). The asymptotic fields \(p^\pm\) and the chemical potential \(\xi^\pm\) transform as follows
\begin{equation}\label{eq:gaugetranp}
    \delta p_\pm = \eta'_\pm \quad \text{and} \quad \delta \xi_\pm = \dot\eta_\pm.
\end{equation}
It can be demonstrated that such transformations are not generated by any constraints; thus, they qualify as symmetry transformations, even though they are not rigid \cite{Banados:1998gg}. Consequently, these transformations yield conserved charges that depend on the choice of the transformation parameters \(\eta^\pm\). Notably, it turns out that the gauge transformation parameters \(\eta^\pm\) are field-dependent. Following the work in \cite{REGGE1974286, Banados:1994tn}, we can compute these conserved charges. It was observed in \cite{Avan:1991kq, Avan:1991ik, Dutta:2023uxe} that there exists an infinite sequence of asymptotic conserved charges represented by
\begin{equation}\label{eq:Qn}
    Q^\pm_n = \pm\frac{\mathrm{k}}{4\pi} \int d\theta \sum_{k=0}^n \frac{^nC_k}{2^k(2k+1)} p^{2k+1}_\pm(t,\theta)W^{n-k}(\theta)
\end{equation}
where \(n=1\) corresponds to the Hamiltonian given in (\ref{eq:HamCFT}). Referring to equation (\ref{eq:poisson}), we see that the Poisson brackets of the charges \(Q^\pm_n\) for all \(n \geq 2\) with the Hamiltonian vanish, indicating that these charges are constants of motion. Additionally, it can be verified that given the Poisson structure described in (\ref{eq:poisson}), the Poisson bracket between any two conserved charges vanishes on-shell
\begin{equation}
  \{Q^\pm_m, Q^\pm_n\}_{PB} = 0 , \quad \forall\ m,n \geq 1.
\end{equation}
This result implies the integrable structure of \(AdS_3\) gravity under the specified boundary conditions.

\subsection{BTZ black hole}

For the chosen Hamiltonian (\ref{eq:HamCFT}), the chemical potentials $\xi^\pm$ can be calculated using the definition given in (\ref{eq:xidef})
\begin{equation}
    \xi^\pm = \mp \left( \frac{p^2_\pm(t,\theta)}{2} + W(\theta) \right).
\end{equation}
In the case of a time-independent geometry, the equations of motion outlined in (\ref{eq:maxeq}) indicate that the chemical potentials $\xi^\pm$ remain constant, and the momenta $p_\pm$ do not depend on time $t$. Consequently, we have the equation
\begin{equation}
    p_\pm(\theta)\partial_\theta p_\pm(\theta) + W'(\theta) = 0.
\end{equation}
This can be expressed equivalently as
\begin{eqnarray}
    \frac{\partial}{\partial\theta} \left( v(\theta)\sigma(\theta) \right) = 0, \quad \text{and} \quad \frac{\partial}{\partial\theta} \left( \frac{\sigma^2(\theta) + v^2(\theta)}{2} \right) + W'(\theta) = 0.
\end{eqnarray}
If \( W'(\theta) \neq 0 \), then \(\sigma(\theta)\) and \(v(\theta)\) are functions of \(\theta\). These solutions correspond to spherically non-symmetric cases. In the scenario where \( W'(\theta) = 0 \), both \( v(\theta) \) and \( \sigma(\theta) \) become constants. After applying appropriate coordinate transformations, we can express the metric in the following form
\begin{equation}\label{eq:BTZmetric}
    ds^2 = - f(r)dt^2 + \frac{dr^2}{f(r)} + r^2 \left( dt + A_\phi d\phi \right)^2 
\end{equation}
where
\begin{eqnarray}
    f(r) = \frac{r^2}{l^2} -\left(\pi ^2 \sigma ^2+v^2 \right) + \frac{A_\phi^2}{r^2}, \quad \tand \quad A_\phi = l \pi v \sigma.
\end{eqnarray}
The metric describes a BTZ black hole characterized by its mass \( M \) and angular momentum \( J \) given by
\begin{equation}\label{eq:MJidentification}
    M = \frac{\pi ^2 \sigma ^2+v^2}{8 G}, \quad \tand \quad J = \frac{l \pi v \sigma}{4G} .
\end{equation}
This indicates that a one-dimensional fluid with constant density and velocity can be associated with a BTZ black hole whose mass and angular momentum are defined by the equations above. For a stable black hole solution to exist, it must satisfy the condition \( J \leq lM \). Consequently, this restricts the fluid velocity, which can be expressed as
\begin{equation}
    |v| \leq \pi \sigma.
\end{equation}

Fluid density \(\sigma\) is always positive, while velocity \(v\) can be either positive or negative. The sign of the fluid velocity determines the direction of the angular momentum of the black hole. Conversely, a BTZ black hole with mass \(M\) and angular momentum \(J\) corresponds to a one-dimensional boundary fluid with constant density and a velocity defined by the following equations
\begin{equation}\label{eq:invMJidentification}
    v \sigma = \frac{4 G J}{\pi  l } \quad \tand \quad \sigma^2 = \frac{4 G \left(\sqrt{l^2
   M^2-J^2}+l M\right)}{\pi ^2 l}.
\end{equation}
The black hole has two horizons, \(r_+\) and \(r_-\), given by
\begin{eqnarray}
    r_+ = l \pi \sigma \quad \tand \quad r_- = |l v|.
\end{eqnarray}
The extremal limit corresponds to \(|v| = \pi \sigma\). A non-rotating BTZ black hole corresponds to \(v = 0\), which indicates a static fluid. The Euclidean temperature of the black hole can be calculated by ensuring there is no conical singularity at \(r = r_+\). The black hole temperature is given by
\begin{equation}
    T = \frac{\pi ^2 \sigma ^2-v^2}{2 \pi
   ^2 l \sigma }.
\end{equation}
Finally, the entropy of the black hole can be expressed as
\begin{equation}
    S = \frac{2\pi r_+}{4G} = \frac{l \pi ^2 \sigma }{2 G}.
\end{equation}
By using the relation outlined in equation (\ref{eq:ppmsv}), we can express all thermodynamic variables in terms of the constant \(p_\pm\). Since, for a black hole solution, \(v \leq \pi \sigma\), \(p_-\) is always negative. The black hole horizon is given by
\begin{eqnarray}
    r_+ = \frac{l}{2}(p_+ + |p_-|).
\end{eqnarray}
Thus, the entropy can also be rewritten as
\begin{equation}\label{eq:BTZentropy}
    S = \frac{\pi l}{4G} \left(p_+ + |p_-|\right).
\end{equation}

Our next objective is to quantize this system, identify the Hilbert space $\cH$, and calculate the degeneracy of the states in $\cH$ that contribute to the black hole entropy for a given set of $M$ and $J$.

%%%%%%%%%%%%%%%%%%%%%%%%%%%%%%%%%%%%%%%%%%%%%%%%%%%%%%%%%%%%%%%%%%%%%%%%%%%%%%%%%%%%%%%%%%%%%%%%%%%%%%%%%%%%%%%%%%%%%%%%%%%%%%%%%%%%%%%%%%%%%%

\section{Quantisation and black hole microstates}
\label{sec:quantisation}

We have observed that for the boundary conditions under consideration, the metric is parameterized by two functions, \( p_\pm(t,\theta) \). These fields adhere to the classical Poisson brackets described in equation (\ref{eq:poisson}). We will now elevate the classical algebras to quantum algebras and construct the corresponding representation of the Hilbert space to examine the properties of the associated quantum metric. Given that the algebras are decoupled for the \(\pm\) sectors, the complete Hilbert space \(\mathcal{H}\) can be expressed as a direct product of the Hilbert spaces of both sectors: \(\mathcal{H} = \mathcal{H}^+ \otimes \mathcal{H}^-\). Our objective is to identify the states within the Hilbert space such that the expectation value of the quantum metric \(d\hat{s}^2\) in those states yields the classical metric. Additionally, we aim to compute the degeneracy of those states to produce the correct black hole entropy.

To quantize the system, we define 
\begin{equation}\label{eq:defptilde}
    p_\pm(t, \theta) = 2\pi\sqrt{\frac{2}{\mathrm{k}}} \tilde p_\pm(t, \theta).
\end{equation}
The Poisson bracket for \(\tilde p_\pm(t, \theta)\) is given by
\begin{equation}\label{eq:ptildepoisson}
    \{\tilde p_\pm(t, \theta), \tilde p_\pm(t, \theta')\}_{PB} = \mp \frac{1}{2\pi} \frac{\partial}{\partial \theta}\delta(\theta - \theta').
\end{equation}
To quantize the system, we need to promote the Poisson bracket to the commutator bracket. This is achieved by multiplying the right-hand side of equation (\ref{eq:ptildepoisson}) by \( i c^{-1} \), where 
\begin{equation}
    c = \frac{3l}{2G}.
\end{equation}
Here, \( c^{-1} \) serves as the equivalent of the Planck constant, and as \( c \) approaches infinity, we revert to the classical limit. This choice of \( c \) plays a pivotal role in determining the number of microstates of BTZ black holes because the spectrum and particularly, the spacing between energy levels is directly influenced by our selection of this parameter. Consequently, the commutation relations between \( \tilde{p}_\pm \) can be expressed as
\begin{equation}
    [\tilde p_\pm(t,\theta), \tilde p_\pm(t, \theta')] = \mp \frac{i}{2\pi c} \frac{\partial}{\partial \theta}\delta(\theta-\theta').
\end{equation}
We express \(\tilde{p}_\pm(t,\theta)\) in terms of its modes (omitting the time dependence of the modes \(\alpha_n^\pm\))
\begin{equation}\label{eq:ptildeexpan}
    \tilde p_\pm(t,\theta) = \frac{1}{2\pi \sqrt{c}} \sum_n \alpha^{\pm}_{\pm n} e^{i n \theta}.
\end{equation}
The modes \(\alpha^\pm_n\) adhere to the following equal-time commutation relation
\begin{equation}\label{eq:balgebra}
    [\alpha^\pm_m , \alpha^\pm_{n}] = m \delta_{m+n}.
\end{equation}

Our next goal is to construct the Hilbert space and establish the correspondence between the states within it and the various $AdS_3$ geometries.

\subsection{Bosonisation and the Hilbert space}

Given that the modes of \(\tilde{p}_\pm(t, \theta)\) correspond to two distinct sets of the Kac-Moody algebra (\ref{eq:balgebra}), we can utilize the bosonization of relativistic Dirac fermions \(\psi_\pm(t, \theta)\). This approach allows us to establish a connection between these fermionic fields and the \(\tilde{p}_\pm(t, \theta)\) fields\footnote{A similar construction of the Hilbert space in terms of bosonic operators was discussed in \cite{Afshar:2017okz,Afshar:2016uax}.}. For clarity and simplicity in the discussion on bosonization, we will temporarily set aside the \(\pm\) signs, as both sectors share an identical analytical framework. However, we will reintroduce them when we delve into the identification of states within the Hilbert space for any specific bulk configuration.

We begin with the theory of free fermions in \( (1+1) \) dimensions. The fermionic fields can be expressed as follows (we omit the time dependence of the modes)
\begin{eqnarray}\label{eq:fermionmodes}
    \psi(t,\theta) = \frac{1}{\sqrt{2\pi}} \sum_{m \in \mathbb{Z}} \psi_{m-1/2} \ e^{i m \theta}, \quad \psi^\dagger(t,\theta) = \frac{1}{\sqrt{2\pi}} \sum_{m\in \mathbb{Z}} \psi^\dagger_{m-1/2} \ e^{-i m \theta}.
\end{eqnarray}
The fermionic modes \( \psi_q \) and \( \psi^\dagger_r \), where \( q, r \in \mathbb{Z} + \frac{1}{2} \), satisfy the following anti-commutation relations
\begin{eqnarray}\label{eq:psialgebra}
    \{\psi_q, \psi_r\} =0, \quad \{\psi^\dagger_q, \psi^\dagger_r\} = 0 \quad \text{and} \quad \{\psi_q, \psi^\dagger_r\} = \delta_{q,r}.
\end{eqnarray}
These relations are derived from the equal-time anti-commutation relations between the fields
\begin{eqnarray}
\begin{split}
   \{\psi(t,\theta),\psi(t,\phi)\} & = 0, \quad \{\psi^\dagger(t,\theta),\psi^\dagger(t,\phi)\} = 0, \\
   \{\psi(t,\theta),\psi^\dagger(t,\phi)\} & = \delta(\theta-\phi).
   \end{split}
\end{eqnarray}

We define the normal ordering as follows
\begin{eqnarray}
    : \psi^\dagger_q \psi_r : = \begin{cases}
       \ \ \psi^\dagger_q \psi_r & r>0 \\
        - \psi_r \psi^\dagger_q & r<0 .
    \end{cases}
\end{eqnarray}
Alternatively, this can be expressed as
\begin{equation}
    : \psi^\dagger_q \psi_r : = \psi^\dagger_q \psi_r - \langle\psi^\dagger_q \psi_r \rangle.
\end{equation}
Next, we introduce the following bosonic operators \cite{Avan:1992gm}
\begin{equation}\label{eq:Boperators}
B^K_n = \sum_{m\in \mathbb{Z}} m^K : \psi^{\dagger}_{m-n-1/2} \psi_{m-1/2}:.
\end{equation}
These bosonic operators satisfy the following commutation relation
\begin{eqnarray}\label{eq:BKcommutation}
\begin{split}
[B^{K_1}_m, B^{K_2}_n]  = \sum_{q\in \mathbb{Z}} & \left( (q-n)^{K_1}q^{K_2} - (q-m)^{K_2}q^{K_1} \right) : \psi^{\dagger}_{q-m-n-1/2} \psi_{q-1/2}: \\
& + \delta_{m+n} \sum_{l=0}^{m-1} l^{K_1}(l-m)^{K_2}.
\end{split}
\end{eqnarray}

We compute the normalized fermion bilinears of \(\psi^\dagger\) and \(\psi\), which is expressed as
\begin{eqnarray}\label{eq:bosonisation}
\begin{split}
  :\psi^\dagger(t,\theta) \psi(t,\theta): & =   \frac{1}{2\pi} \sum_{m,n \in \mathbb{Z}} :\psi^\dagger_{m-n-1/2} \psi_{m-1/2}: e^{i n \theta} = \frac{1}{2\pi} \sum_{n\in \mathbb{Z}} B^0_n e^{i n \theta}.
\end{split}
\end{eqnarray}
Here, \(B^0_n\) satisfies the Kac-Moody algebra
\begin{eqnarray}
    [B^0_m, B^0_n] = m \delta_{m+n}.
\end{eqnarray}
We identify \(B^0_n\) as the modes of the bosonic fields \(\tilde{p}(t,\theta)\) : \(B^0_n \equiv \alpha_n\). Therefore, from equation (\ref{eq:ptildeexpan}), we assert that the fermion bilinear can be precisely expressed as
\begin{equation}\label{eq:bosonisation2}
	:\psi^\dagger(t,\theta) \psi(t,\theta): \ \equiv \sqrt{c}\  \tilde{p}_\pm(t,\theta).
\end{equation}
This demonstrates that collective fermionic excitations \(:\psi^\dagger(t,\theta) \psi(t,\theta):\) indeed behave like bosonic excitations, a fundamental aspect of bosonization\footnote{See \cite{Dhar:1992hr,Dhar:2005qh, Das:1995gd, Das:2004rx, Rao:2000rh, Miranda:2003ga} for bosonization in similar context and some review on the topic.}. The corresponding inverse relation is represented by
\begin{equation}
    \psi(\theta) = \frac{1}{\sqrt{2\pi}} : e^{-2\pi i \varphi(\theta)} :
\end{equation}
where
\begin{equation}
    \tilde p(\theta) =  \frac{ \varphi'(\theta)}{\sqrt{c}}.
\end{equation}
Higher-order bilinears can also be expressed as higher powers of the bosonized field \(\tilde{p}(\theta)\). For example, consider the following
\begin{eqnarray}\label{eq:ptildesquare}
\begin{split}
   2\pi c :\frac{\tilde p^2(\theta)}{2}: \ & = -\frac{i}{2} :\left( \psi^\dagger \partial_\theta \psi - \partial_\theta  \psi^\dagger \psi\right) : \\
   & =  \frac{1}{2\pi} \sum_p \left(B_p^1 - \frac{p}{2} B_p^0 \right)e^{i p \theta}.
   \end{split}
\end{eqnarray}
In addition, we further find that
\begin{eqnarray}
\begin{split}
   (2\pi)^2 c^{3/2} :\frac{\tilde p^3(\theta)}{3}:\ 
    & = -\frac{1}{6} :\partial_{\theta}^2\psi^\dagger\psi + \psi^\dagger \partial_\theta^2\psi - 4 \partial_\theta\psi^\dagger\partial_\theta\psi: \\
    & =  \frac{1}{2\pi} \sum_p \lb B^2_p - p B^1_p + \frac{p^2}{6} B^0_p\rb e^{i p \theta}.
    \end{split}
\end{eqnarray}

%%%%%%%%%%%%%%%%%%%%%%%%%%%%%%%%%%%%%%%%%%%%%%
%%%%%%%%%%%%%%%%%%%%%%%%%%%%%%%%%%%%%%%%%%%%%%
%%%%%%%%%%%%%%%%%%%%%%%%%%%%%%%%%%%%%%%%%%%%%%

\subsection{The Hilbert space}

\begin{figure}[h]
	\centering
	\begin{subfigure}[b]{0.35\textwidth}
		\centering
		\includegraphics[width=0.35\textwidth]{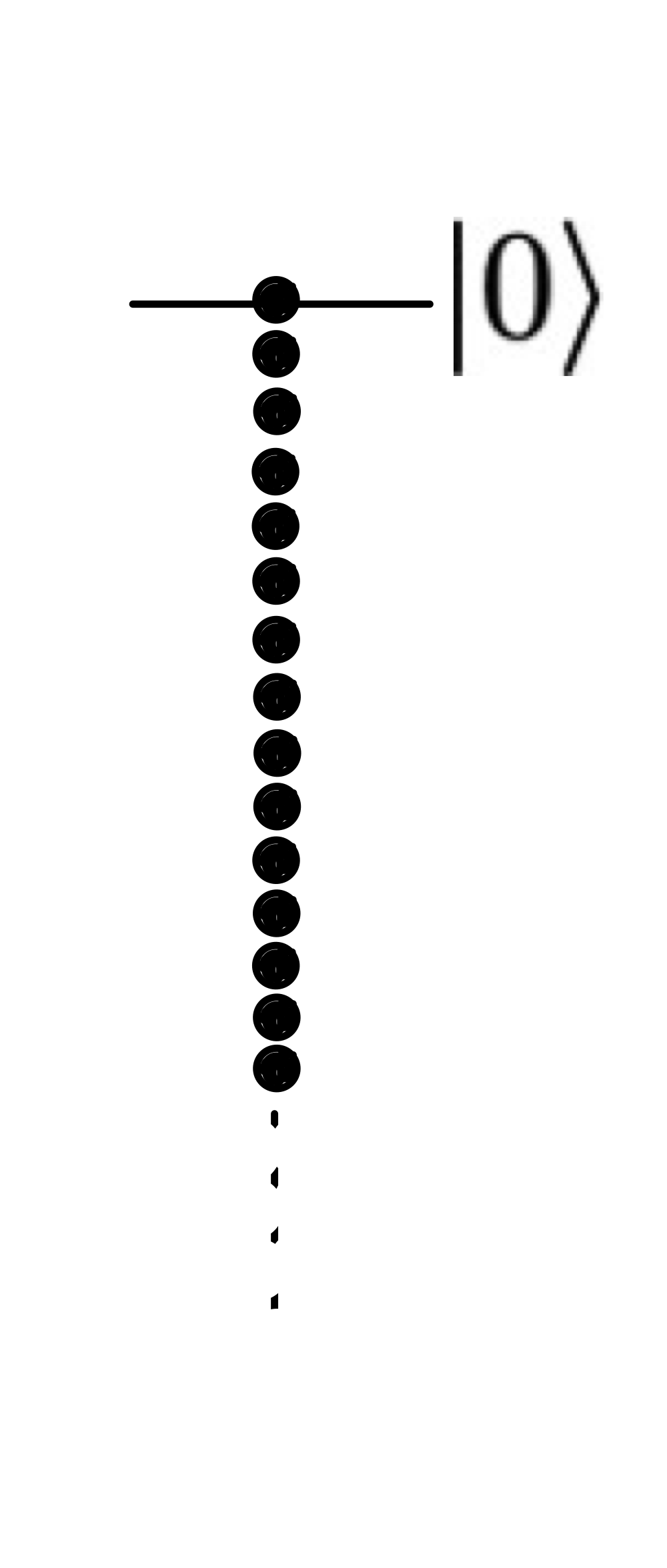}
		\caption{}
		\label{fig:absolutegs}
	\end{subfigure}
	\
	\begin{subfigure}[b]{0.5\textwidth}
		\centering
		\includegraphics[width=0.5\textwidth]{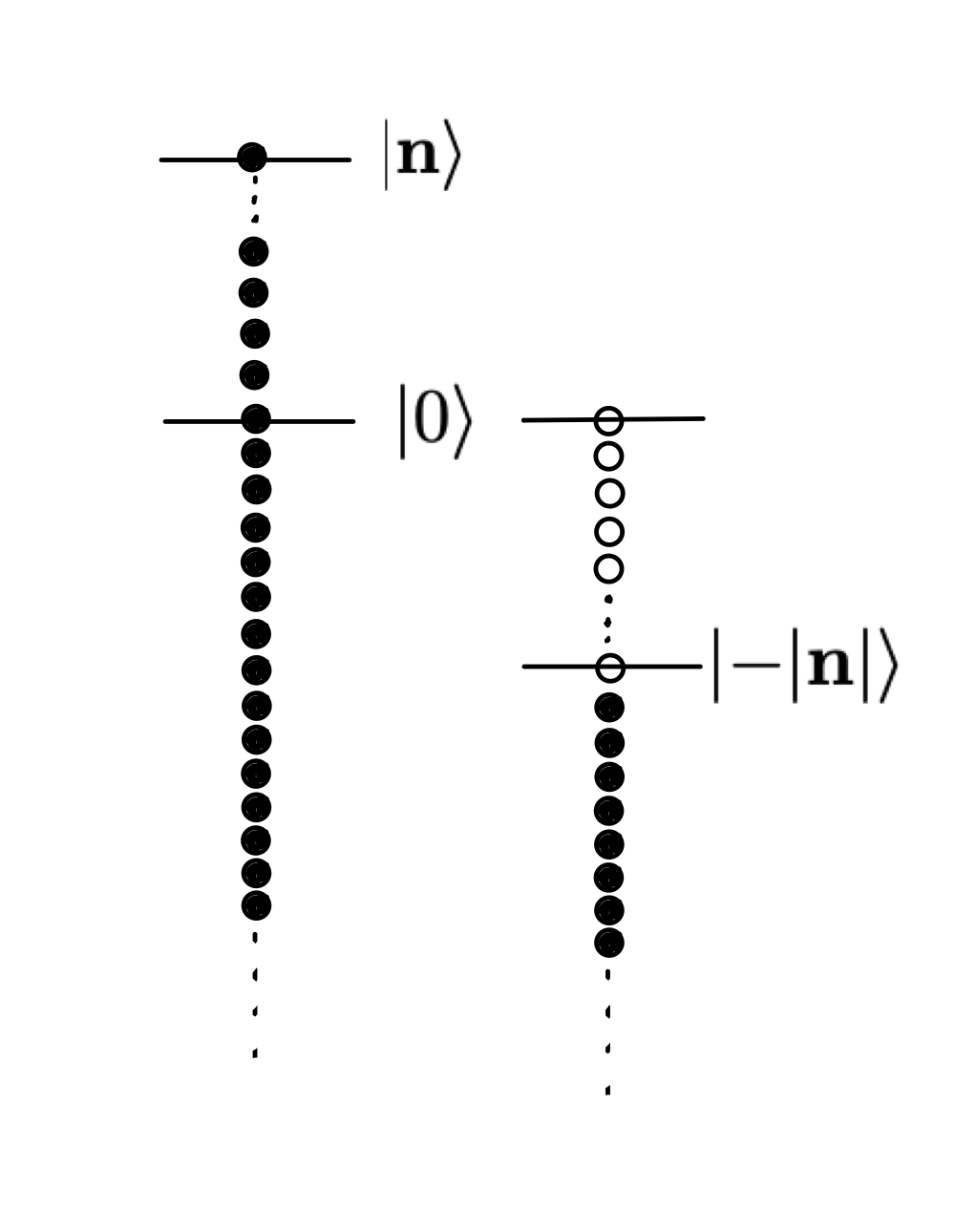}
		\caption{}
		\label{fig:npstate}
	\end{subfigure}
	\caption{(a) Absolute ground state $\ket{0}$.
   (b) A $n$-particles for $\ket{\mathbf{n}>0}$ and $\ket{\mathbf{n}<0}$ (these are hole states).}
	\label{fig:gsandnps}
\end{figure}

To find a representation of the algebra (\ref{eq:psialgebra}), we define the Dirac vacuum \(\ket{0}\) by the following relations
\begin{eqnarray}
   \psi_{k+\frac12}\ket{0} & = & 0, \quad \forall \quad k \geq 0\\
    \psi^\dagger_{k-\frac12}\ket{0} & = & 0, \quad \forall \quad k \leq 0 .
\end{eqnarray}
The ground state can be thought of as the "Dirac sea," which occupies all the states up to \(k = 0\) (see Fig. \ref{fig:absolutegs}). Excited states can be generated by applying any number of \(\psi^\dagger_{q}\) for \(q > 0\) and \(\psi_{q}\) for \(q < 0\). The state \(\psi^\dagger_{q > 0} \ket{0}\) can be interpreted as a single fermion state above the Dirac sea, whereas \(\psi_{q < 0} \ket{0}\) represents a hole state, signifying the absence of a fermion from the Dirac sea.
Any generic state with $n_p$ particles and $n_h$ holes excited above the absolute ground state $\ket{0}$ can be expressed as:
\begin{eqnarray}
	\prod_{i=0}^{n_p} \prod_{j=0}^{n_h} \psi^\dagger_{n_i +\frac12} \psi_{-m_j -\frac12}\ket{0}.
\end{eqnarray}

We define an \(\mathbf{n}\)-particle normalized ground state as follows (see Fig. \ref{fig:npstate})
\begin{equation}\label{eq:nstate}
\ket{\mathbf{n}} =
\begin{cases}
       \psi^\dagger_{\mathbf{n} - \frac12} \cdots \psi^\dagger_{\frac12}\ket{0} & \mathbf{n} >0 \\
        \psi_{\mathbf{n} + \frac12} \cdots \psi_{-\frac12}\ket{0} & \mathbf{n} < 0.
    \end{cases}
\end{equation}
For \(\mathbf{n} < 0\), these represent hole states, meaning particles that have been removed from the Dirac sea. Using the commutation relations given by
\begin{eqnarray}\label{eq:Bpsicommutator}
	[B^K_m, \psi_{k-\frac12}] = - (k+m)^K\psi_{k + m -\frac12}, \quad [B^K_m, \psi^\dagger_{k-\frac12}] = (k-m)^K\psi^\dagger_{k - m -\frac12}
\end{eqnarray}
we can show that
\begin{eqnarray}
    B^0_0 \ket{\mathbf{n}} = \mathbf{n} \ket{\mathbf{n}}.
\end{eqnarray}
Additionally, we find that
\begin{equation}
    B^0_m \ket{\mathbf{n}} =0, \quad \forall \ m>0 \ \text{and} \ \mathbf{n}.
\end{equation}
The \(\mathbf{n}\)-particle Hilbert space \(\cH_\mathbf{n}\) is generated by all possible particle-hole excitations on the corresponding \(\mathbf{n}\)-particle ground state \(\ket{\mathbf{n}}\). It can be demonstrated that \cite{Haldane:1981zza} the bosonic excitations given by
\begin{equation}\label{eq:kstate}
    \ket{\vec k; {\mathbf{n}}} \equiv \ket{k_1, k_2, \cdots} = \prod_{m>0} \left( B^0_{-m}\right)^{k_m} \ket{\mathbf{n}}
\end{equation}
spans $\cH_{\mathbf{n}}$. For example consider the 2-particle ground state $\ket{2} = \psi^\dagger_{\frac32}\psi^\dagger_\frac12\ket{0}$. Applying the operator \(B^0_{-3}\) to \(\ket{2}\) gives: $B^0_{-3} \ket{2} = \psi^\dagger_{\frac52}\psi_{-\frac12} \ket{2} - \psi^\dagger_{\frac72}\psi^\dagger_{\frac32}\ket{0} + \psi^\dagger_{\frac92} \psi^\dagger_{\frac12} \ket{0}$ which represents a superposition of two 2-particle states and a particle-hole pair excitation over the 2-particle ground state. Conversely, a 1-particle excited state \(\psi^\dagger_{\frac{5}{2}} \ket{0}\) can be expressed as \((B^0_{-1})^2 \ket{1}\). Since \(B^0_0\) commutes with \(B^0_m\), the states \(\ket{\vec{k}}\) are eigenstates of \(B^0_0\) with eigenvalue \(\mathbf{n}\). The basis vectors satisfy the normalization condition
\begin{equation}\label{eq:knormalisation}
    \braket{\vec{k};\mathbf{n}}{\vec{k}';\mathbf{n}} = z_{\vec{k}}\delta_{\vec{k},\vec{k}'}
\end{equation}
where
\begin{equation}
    z_{\vec{k}} = \prod_m k_m! m^{k_m}.
\end{equation}

Thus, we conclude that the fermionic Hilbert space \(\cH\) can be decomposed into
\begin{equation}
    \cH = \cH_0 \oplus \cH_{1} \oplus \cH_2 \oplus \cdots.
\end{equation}

There is another way to express the excited states in \(\cH_{\mathbf{n}}\) that will be relevant for us. We can excite any number of particle-hole pairs above the \(\mathbf{n}\)-particle ground state. For \(\mathbf{n} > 0\), these states can be written as follows
\begin{equation}\label{eq:mnstate}
    \ket{\vec{m}; \vec{n};\mathbf{n}} = \prod_{i} \psi^\dagger_{\mathbf{n}+n_i +\frac12} \psi_{\mathbf{n}- m_i-\frac12}\ket{\mathbf{n}},\quad \forall\ n_i, m_i \geq 0.
\end{equation}
Each pair \(\psi^\dagger_{\mathbf{n}+n_i +\frac{1}{2}} \psi_{\mathbf{n}-m_i-\frac{1}{2}}\) creates a hole at or below \(k = \mathbf{n}\) and a particle above \(k = \mathbf{n}\). We can represent these states in terms of Young diagrams \(\ket{R}\) (up to an overall sign) of irreducible \(U(\infty)\) representations \cite{Marino:2005sj}. If the lengths of the \(i^{\text{th}}\) row and column of a diagram are given by \(l_i\) and \(v_i\), then we have \(m_i = v_i - i\) and \(n_i = l_i - i\). Thus, we can denote a generic excited state using a Young diagram
\begin{eqnarray}
    \ket{R;\mathbf{n}} \equiv \ket{\vec{m}; \vec{n};\mathbf{n}}.
\end{eqnarray}
The total number of boxes in such diagrams is given by \(\sum_i(n_i+m_i+1)\). States with distinct \(\vec{m}\) and \(\vec{n}\), portrayed by different Young diagrams, correspond to different particle-hole excitations above \(\ket{\mathbf{n}}\) and are orthonormal
\begin{equation}
    \braket{R'; \mathbf{n}}{R;\mathbf{n}} = \delta_{R'R}.
\end{equation}
Moreover, these excited states can be expressed as a linear superposition of \(\ket{\vec{k}; \mathbf{n}}\), as defined in \((\ref{eq:kstate})\).

Similarly, we can express the excited states for \(\mathbf{n} < 0\) as
\begin{equation}\label{eq:mnstate-}
    \ket{\vec{m}; \vec{n};\mathbf{n}} = \ket{R;\mathbf{n}} = \prod_{i} \psi^\dagger_{\mathbf{n} + n_i + \frac12} \psi_{\mathbf{n}-m_i-\frac12}\ket{\mathbf{n}},\quad \forall\ n_i, m_i \geq 0.
\end{equation}
These states also represent particle-hole excitations above the ground state \(\ket{\mathbf{n}}\) and can be represented by Young diagrams in a similar way.

\subsection{States in the Hilbert space and $AdS_3$ geometry}

In the Chern-Simons formulation of gravity, any bulk configuration is determined by the choice of the boundary Hamiltonians. In Section \ref{sec:colHamBTZ}, we consider the boundary Hamiltonian as the collective field theory Hamiltonian and determine that this boundary condition permits a BTZ black hole solution in the bulk. Since the dynamics of the collective field can be equated to a one-dimensional fluid moving on a circle, a BTZ black hole with mass \(M\) and angular momentum \(J\) corresponds to a one-dimensional fluid with constant density moving at a constant velocity (as shown in equation \ref{eq:MJidentification}).

In the quantum picture, we first elevate all classical variables \(\sigma(t, \theta)\) and \(v(t, \theta)\) (or, alternatively, \(p_\pm(t, \theta)\)) to operators in the Hilbert space. A classical configuration, represented by equation (\ref{eq:BTZmetric}), corresponds to a state \(\ket{\Psi} = \ket{\Psi^+} \otimes \ket{\Psi^-}\) in the Hilbert space \(\cH = \cH^+ \otimes \cH^-\), such that
\begin{equation}
    \bra{\Psi}:\hat v(t,\theta) \hat \sigma(t,\theta):\ket{\Psi} = \frac{4GJ}{\pi l} \quad \tand \quad \bra{\Psi}:\pi^2 \hat \sigma^2(t,\theta) + v^2(t,\theta):\ket{\Psi} = 8GM.
\end{equation}
Alternatively, this can be expressed as
\begin{equation}\label{eq:pMJrelation}
    \bra{\Psi^\pm} : \hat p_\pm^2(t,\theta) : \ket{\Psi^\pm} = \frac{12}{c}\left( M l \pm J \right).
\end{equation}

For a given state $\ket{\Psi^\pm} \in \cH^\pm$, there is a unique bulk geometry. However, the converse may not hold true, which leads to degeneracy.

By utilizing the bosonization relation (\ref{eq:ptildesquare}), we find that
\begin{eqnarray}
    :\hat p^2_\pm(t,\theta): = \frac{24}{c^2} \sum_p \left(B^1_p -\frac{p}{2}B_p^0 \right) e^{i p \theta}.
\end{eqnarray}
The boundary Hamiltonian (\ref{eq:HamCFT}) can also be expressed as
\begin{eqnarray}
    H^\pm = \pm \frac{\sqrt{3}}{c^2} B^2_0.
\end{eqnarray}
All other conserved charges (\ref{eq:Qn}) (for \( W(\theta) = 0 \)) can be written as
\begin{equation}
    Q^\pm_n = \pm \frac{1}{2\sqrt{3}}\left( \frac{6}{c^2}\right)^n B^{2n}_0.
\end{equation}
Using (\ref{eq:BKcommutation}), one can easily verify that all the conserved charges \( Q^\pm_m \) commute with each other.

Since these conserved charges correspond to the symmetries of the boundary theory, the ground state must be annihilated by these charges. This implies that the \emph{zero} particle state \( \ket{0^+} \otimes \ket{0^-} \) is the required ground state of \( \mathcal{H} \). This state is associated with a massless BTZ black hole solution because
\begin{equation}
    \bra{0^\pm}:\hat p_\pm^2(t,\theta):\ket{0^\pm} = 0.
\end{equation}

We now consider a generic excited state \(\ket{R^\pm;0} \equiv \ket{R^\pm}\) in the zero-particle Hilbert space \(\mathcal{H}_0^\pm\). By using the commutation relations given in equation (\ref{eq:Bpsicommutator}), we can show that
\begin{equation}
    \bra{R^\pm}B^1_p\ket{R^\pm} = |R^\pm| \delta_{p,0}, \quad \bra{R^\pm}B^0_p\ket{R^\pm} = 0
\end{equation}
where \(|R^\pm| = m_i+n_i+1\) represents the number of boxes in the Young diagram \(R^\pm\). Thus, we have the following relationship
\begin{equation}\label{eq:pRrelation}
    \bra{R^\pm}:\hat p^2_\pm(t,\theta): \ket{R^\pm} = \frac{24}{c^2} |R^\pm|.
\end{equation}
By comparing equations (\ref{eq:pMJrelation}) and (\ref{eq:pRrelation}), we find that
\begin{equation}\label{eq:boxnumberMJrel}
    |R^\pm| = \frac{c}{2} \left( M l \pm J \right).
\end{equation}
The Young diagram states \(\ket{R^\pm}\), which contain \(|R^\pm|\) boxes, represent the microstates of a BTZ black hole with mass \(M\) and angular momentum \(J\). From equation (\ref{eq:gaugetranp}), we observe that for the asymptotic conserved charges (\ref{eq:Qn}), the variation \(\delta p_\pm\) is zero for a static on-shell configuration. In the corresponding quantum framework, we find that the various microstates of a BTZ black hole are the eigenstates of the conserved charges \(Q^\pm_n\) with eigenvalues depending on different higher Casimirs of $R^\pm$ representations.

In the classical limit, which corresponds to the large \(c\) limit, we see that the states corresponding to a BTZ black hole have a large number of boxes. Consequently, all Young diagrams with a total number of boxes given by (\ref{eq:boxnumberMJrel}) represent the microstates for a BTZ black hole with mass \(M\) and angular momentum \(J\). From (\ref{eq:pRrelation}), we can conclude that the classical value of \(p_\pm\) is given by
\begin{equation}
    p^2_\pm = \bra{R^\pm}:\hat p^2_\pm(t,\theta):\ket{R^\pm} = \frac{24}{c^2} |R^\pm|.
\end{equation}
In the large \(c\) limit, we can calculate the degeneracy of the classical geometry given by \(p^2_\pm\). This is expressed by the Hardy-Ramanujan formula
\begin{eqnarray}
    d(p_+, p_-) = e^{2\pi \sqrt{\frac{|R^+|}{6}}+2\pi \sqrt{\frac{|R^-|}{6}}}.
\end{eqnarray}
As a result, the statistical entropy is given by
\begin{equation}
    S_{\text{stat}} = \ln d(p_+, p_-) = 2\pi \sqrt{\frac{|R^+|}{6}}+2\pi \sqrt{\frac{|R^-|}{6}} = \frac{\pi l}{4G} \left(p_+ + |p_-|\right)
\end{equation}
which matches the thermodynamic entropy given in equation (\ref{eq:BTZentropy}).

\section{Conclusion}

In this paper, we study the boundary conditions for spin-two fields in \(AdS_3\), where the boundary dynamics are governed by the interacting collective field theory Hamiltonian proposed by Jevicki and Sakita. Under these boundary conditions, the chemical potentials specifically, the time components of the gauge fields are not constant but depend on the field components in the \(\theta\) direction. Furthermore, it was shown in \cite{Dutta:2023uxe} that the system admits an infinite sequence of conserved charges and reveals an integrable structure.

We quantize this system by promoting the classical Poisson structure to commutator brackets, and we find that the asymptotic symmetry algebra is isomorphic to the \(U(1)\) Kac-Moody algebra, without any central charge. Utilizing quantum bosonization of relativistic free fermions, we relate these fermions to the dynamical fields of \(AdS_3\) gravity.

We construct the Hilbert space of the free fermionic system and link the states in this Hilbert space to the bulk geometries. We discover that different microstates can be represented as irreducible Young diagrams corresponding to \(U(\infty)\) representations, with the number of boxes depending on the mass and angular momentum of the BTZ black hole. In \cite{Perez:2016vqo, Melnikov:2018fhb, Ojeda:2019xih}, the asymptotic growth of microstates for similar types of boundary symmetries was discussed, assuming the existence of a two-dimensional field theory with Lifshitz scaling. In this work, we explicitly identify the allowed quantum states (microstates) for a given bulk configuration, particularly for the BTZ black hole geometry.

Our analysis in this paper serves as a foundation for exploring higher-spin field dynamics. It would be interesting to extend our construction to higher spin fields as well. It was observed in \cite{Dutta:2023uxe} that one can include the classical dynamics of higher spin fields for the same boundary condition. In such a scenario, we may have to extend the algebra of bosonic operators (\ref{eq:BKcommutation}) to include other higher spin operators following \cite{Avan:1992gm}.

\vspace{1cm}
\paragraph{Acknowledgment} We thank Nabamita Banerjee, Ranveer Singh for useful discussions. We are indebted to people of India for their unconditional support toward the researches in basic science.

\bibliographystyle{hieeetr}
\bibliography{ads}{}

\begin{thebibliography}{10}

\bibitem{Brown:1986nw}
J.~D. Brown and M.~Henneaux, ``{Central Charges in the Canonical Realization of Asymptotic Symmetries: An Example from Three-Dimensional Gravity},'' {\em Commun. Math. Phys.}, vol.~104, pp.~207--226, 1986.

\bibitem{Henneaux:2013dra}
M.~Henneaux, A.~Perez, D.~Tempo, and R.~Troncoso, ``{Chemical potentials in three-dimensional higher spin anti-de Sitter gravity},'' {\em JHEP}, vol.~12, p.~048, 2013, 1309.4362.

\bibitem{Bunster:2014mua}
C.~Bunster, M.~Henneaux, A.~Perez, D.~Tempo, and R.~Troncoso, ``{Generalized Black Holes in Three-dimensional Spacetime},'' {\em JHEP}, vol.~05, p.~031, 2014, 1404.3305.

\bibitem{Perez:2016vqo}
A.~P\'erez, D.~Tempo, and R.~Troncoso, ``{Boundary conditions for General Relativity on AdS$_{3}$ and the KdV hierarchy},'' {\em JHEP}, vol.~06, p.~103, 2016, 1605.04490.

\bibitem{Perez:2012cf}
A.~Perez, D.~Tempo, and R.~Troncoso, ``{Higher spin gravity in 3D: Black holes, global charges and thermodynamics},'' {\em Phys. Lett. B}, vol.~726, pp.~444--449, 2013, 1207.2844.

\bibitem{Afshar:2016wfy}
H.~Afshar, S.~Detournay, D.~Grumiller, W.~Merbis, A.~Perez, D.~Tempo, and R.~Troncoso, ``{Soft Heisenberg hair on black holes in three dimensions},'' {\em Phys. Rev. D}, vol.~93, no.~10, p.~101503, 2016, 1603.04824.

\bibitem{Grumiller:2016pqb}
D.~Grumiller and M.~Riegler, ``{Most general AdS$_{3}$ boundary conditions},'' {\em JHEP}, vol.~10, p.~023, 2016, 1608.01308.

\bibitem{Grumiller:2016kcp}
D.~Grumiller, A.~Perez, S.~Prohazka, D.~Tempo, and R.~Troncoso, ``{Higher Spin Black Holes with Soft Hair},'' {\em JHEP}, vol.~10, p.~119, 2016, 1607.05360.

\bibitem{Grumiller:2019tyl}
D.~Grumiller and W.~Merbis, ``{Near horizon dynamics of three dimensional black holes},'' {\em SciPost Phys.}, vol.~8, no.~1, p.~010, 2020, 1906.10694.

\bibitem{Afshar:2016kjj}
H.~Afshar, D.~Grumiller, W.~Merbis, A.~Perez, D.~Tempo, and R.~Troncoso, ``{Soft hairy horizons in three spacetime dimensions},'' {\em Phys. Rev. D}, vol.~95, no.~10, p.~106005, 2017, 1611.09783.

\bibitem{Afshar:2013vka}
H.~Afshar, A.~Bagchi, R.~Fareghbal, D.~Grumiller, and J.~Rosseel, ``{Spin-3 Gravity in Three-Dimensional Flat Space},'' {\em Phys. Rev. Lett.}, vol.~111, no.~12, p.~121603, 2013, 1307.4768.

\bibitem{Troessaert:2013fma}
C.~Troessaert, ``{Enhanced asymptotic symmetry algebra of $AdS_{3}$},'' {\em JHEP}, vol.~08, p.~044, 2013, 1303.3296.

\bibitem{Avery:2013dja}
S.~G. Avery, R.~R. Poojary, and N.~V. Suryanarayana, ``{An sl(2,$\mathbb{R}$) current algebra from $AdS_3$ gravity},'' {\em JHEP}, vol.~01, p.~144, 2014, 1304.4252.

\bibitem{Ammon:2017vwt}
M.~Ammon, D.~Grumiller, S.~Prohazka, M.~Riegler, and R.~Wutte, ``{Higher-Spin Flat Space Cosmologies with Soft Hair},'' {\em JHEP}, vol.~05, p.~031, 2017, 1703.02594.

\bibitem{Ozer:2019nkv}
H.~T. \"Ozer and A.~Filiz, ``{Exploring new boundary conditions for $\mathcal {N}=(1,1)$ extended higher-spin $AdS_3$ supergravity},'' {\em Eur. Phys. J. C}, vol.~80, no.~11, p.~1072, 2020, 1907.06104.

\bibitem{Gonzalez:2018jgp}
H.~A. Gonz\'alez, J.~Matulich, M.~Pino, and R.~Troncoso, ``{Revisiting the asymptotic dynamics of General Relativity on AdS$_{3}$},'' {\em JHEP}, vol.~12, p.~115, 2018, 1809.02749.

\bibitem{Campoleoni:2010zq}
A.~Campoleoni, S.~Fredenhagen, S.~Pfenninger, and S.~Theisen, ``{Asymptotic symmetries of three-dimensional gravity coupled to higher-spin fields},'' {\em JHEP}, vol.~11, p.~007, 2010, 1008.4744.

\bibitem{Dutta:2023uxe}
S.~Dutta, D.~Mukherjee, and S.~Parihar, ``{Higher-spin gravity in AdS3 and folds on Fermi surface},'' {\em Phys. Rev. D}, vol.~107, no.~10, p.~106010, 2023, 2302.08471.

\bibitem{sakita}
A.~Jevicki and B.~Sakita, ``{The quantum collective field method and its application to the planar limit},'' {\em Nuclear Physics B}, vol.~165, no.~3, pp.~511 -- 527, 1980.

\bibitem{jevicki}
A.~Jevicki and B.~Sakita, ``{Collective field approach to the large-$N$ limit: Euclidean field theories},'' {\em Nuclear Physics B}, vol.~185, no.~1, pp.~89 -- 100, 1981.

\bibitem{Campoleoni:2011hg}
A.~Campoleoni, S.~Fredenhagen, and S.~Pfenninger, ``{Asymptotic W-symmetries in three-dimensional higher-spin gauge theories},'' {\em JHEP}, vol.~09, p.~113, 2011, 1107.0290.

\bibitem{Banados:1998gg}
M.~Banados, ``{Three-dimensional quantum geometry and black holes},'' {\em AIP Conf. Proc.}, vol.~484, no.~1, pp.~147--169, 1999, hep-th/9901148.

\bibitem{REGGE1974286}
T.~Regge and C.~Teitelboim, ``Role of surface integrals in the hamiltonian formulation of general relativity,'' {\em Annals of Physics}, vol.~88, no.~1, pp.~286--318, 1974.

\bibitem{Banados:1994tn}
M.~Banados, ``{Global charges in Chern-Simons field theory and the (2+1) black hole},'' {\em Phys. Rev. D}, vol.~52, pp.~5816--5825, 1996, hep-th/9405171.

\bibitem{Avan:1991kq}
J.~Avan and A.~Jevicki, ``{Classical integrability and higher symmetries of collective string field theory},'' {\em Phys. Lett. B}, vol.~266, pp.~35--41, 1991.

\bibitem{Avan:1991ik}
J.~Avan and A.~Jevicki, ``{Quantum integrability and exact eigenstates of the collective string field theory},'' {\em Phys. Lett. B}, vol.~272, pp.~17--24, 1991.

\bibitem{Afshar:2017okz}
H.~Afshar, D.~Grumiller, M.~M. Sheikh-Jabbari, and H.~Yavartanoo, ``{Horizon fluff, semi-classical black hole microstates \textemdash{} Log-corrections to BTZ entropy and black hole/particle correspondence},'' {\em JHEP}, vol.~08, p.~087, 2017, 1705.06257.

\bibitem{Afshar:2016uax}
H.~Afshar, D.~Grumiller, and M.~M. Sheikh-Jabbari, ``{Near horizon soft hair as microstates of three dimensional black holes},'' {\em Phys. Rev. D}, vol.~96, no.~8, p.~084032, 2017, 1607.00009.

\bibitem{Avan:1992gm}
J.~Avan and A.~Jevicki, ``{Interacting theory of collective and topological fields in two-dimensions},'' {\em Nucl. Phys. B}, vol.~397, pp.~672--704, 1993, hep-th/9209036.

\bibitem{Dhar:1992hr}
A.~Dhar, G.~Mandal, and S.~R. Wadia, ``{Nonrelativistic fermions, coadjoint orbits of $W_\infty$ and string field theory at $c = 1$},'' {\em Mod. Phys. Lett.}, vol.~A7, pp.~3129--3146, 1992, hep-th/9207011.

\bibitem{Dhar:2005qh}
A.~Dhar, ``{Bosonization of non-relativistic fermions in 2-dimensions and collective field theory},'' {\em JHEP}, vol.~07, p.~064, 2005, hep-th/0505084.

\bibitem{Das:1995gd}
S.~R. Das and S.~D. Mathur, ``{Folds, bosonization and nontriviality of the classical limit of 2-D string theory},'' {\em Phys. Lett. B}, vol.~365, pp.~79--86, 1996, hep-th/9507141.

\bibitem{Das:2004rx}
S.~R. Das, ``{D-branes in 2-d string theory and classical limits},'' in {\em {3rd International Symposium on Quantum Theory and Symmetries}}, pp.~218--233, 1 2004, hep-th/0401067.

\bibitem{Rao:2000rh}
S.~Rao and D.~Sen, ``{An Introduction to bosonization and some of its applications},'' 5 2000, cond-mat/0005492.

\bibitem{Miranda:2003ga}
E.~Miranda, ``{Introduction to bosonization},'' {\em Braz. J. Phys.}, vol.~33, pp.~3--35, 2003.

\bibitem{Haldane:1981zza}
F.~D.~M. Haldane, ``{Luttinger liquid theory of one-dimensional quantum fluids. I. Properties of the Luttinger model and their extension to the general 1D interacting spinless Fermi gas},'' {\em J. Phys. C}, vol.~14, pp.~2585--2609, 1981.

\bibitem{Marino:2005sj}
M.~Marino, ``{Chern-Simons theory, matrix models, and topological strings},'' {\em Int. Ser. Monogr. Phys.}, vol.~131, pp.~1--197, 2005.

\bibitem{Melnikov:2018fhb}
D.~Melnikov, F.~Novaes, A.~P\'erez, and R.~Troncoso, ``{Lifshitz Scaling, Microstate Counting from Number Theory and Black Hole Entropy},'' {\em JHEP}, vol.~06, p.~054, 2019, 1808.04034.

\bibitem{Ojeda:2019xih}
E.~Ojeda and A.~P\'erez, ``{Boundary conditions for General Relativity in three-dimensional spacetimes, integrable systems and the KdV/mKdV hierarchies},'' {\em JHEP}, vol.~08, p.~079, 2019, 1906.11226.

\end{thebibliography}

\end{document}